# Solitary pulmonary nodules prediction for lung cancer patients using nomogram and machine learning

Hailan Zhang[1] Gongjin Song[1]

[1]School of Computer Science and Technology, Anhui University of Technology, Maanshan, China

## Abstract

***Background:*** Lung cancer (LC) is a type of malignant neoplasm that originates in the bronchial mucosa or glands. As a clinically common nodule, solitary pulmonary nodules (SPNs) have a significantly higher probability of malignancy when they are larger than 8 mm in diameter. But there is also a risk of lung cancer when the diameter is less than 8mm, the purpose of this study was to create a nomogram for estimating the likelihood of lung cancer in patients with SPNs of 8 mm or smaller using computed tomography (CT) scans and biomarker information.

***Objective.*** Use CT scans and various biomarkers as input to build predictive models for the likelihood of lung cancer in patients with SPNs of 8 mm or less.

***Results:*** The age, precursor gastrin-releasing peptide (ProGRP), Carcinoembryonic Antigen (CEA), gender and stress corrosion cracking (SCC) were independent key tumor markers and were entered into the nomogram. The developed nomogram demonstrated strong accuracy in predicting lung cancer risk, with an internal validation area under the receiver operating characteristics curve (ROC) of 0.8474. The calibration curves plotted showed that the nomogram predicted the probability of lung cancer with good agreement with the actual probability.

***Conclusions:*** In this study, we finally succeeded in constructing a suitable nomogram that could predict the risk of lung cancer in patients with SPNs ≤8 mm in diameter. The model has a high level of accuracy and is able to accurately distinguish between different patients, allowing clinicians to develop personalized treatment plans for individuals with SPNs.

# 1. Introduction

The data show that as a popular cancer, lung cancer causes more than 2.2 million new cases and more than 1.8 million deaths in 2020 [1-3]. Lung cancer far exceeds all other cancer types regarding the number of patients and deaths, ranking first in the death rate [4,5].

Solitary pulmonary nodules (SPNs) are defined as single, round, lung lesions that are 30 millimeters or smaller in diameter and are not accompanied by atelectasis, pleural effusion, or other related symptoms [6]. According to the size of the node diameter, SPNs can be divided into two types with 8mm as the boundary, those smaller than 8mm are recorded as small nodules, and those larger than 8mm are typical solitary pulmonary nodules [7]. SPNs can be considered as one of the early signs of lung cancer. Some pulmonary nodules may transform into malignant tumors. Treatment of patients with SPNs can significantly reduce the incidence of lung cancer [8].

Currently, it is difficult to accurately determine whether SPNs will become malignant tumors without performing a histological evaluation using frozen tissue samples [9]. In some places with poor medical conditions, in order to ensure patients' safety, clinicians usually treat patients with SPNs in a one-size-fits-all treatment method, that is, direct resection of the diseased part. Therefore, it is necessary to construct a model that effectively predicts whether SPNs will transform into malignant tumors [10].

Nomograms are commonly used as simple, statistical tools to predict the likelihood of a particular event, such as cancer. They are often used in a clinical setting to help quantify the probability of a particular outcome [11]. The nomogram can predict whether SPNs will transform into malignant tumors by computing the score for patients with SPNs, satisfying our desire for clinically integrated models and our drive for personalized medicine [12].

## 2. Patients and Methods

*2.1. Ethics statement.* The study followed the guidelines of the Declaration of Helsinki and verbal informed consent was obtained from all subjects by telephone after the study had been fully explained to them.

*2.2. Patients selection.* Retrospective analysis of the clinical data of 421 patients with SPNs who underwent surgical resection at CHAOHU HOSPITAL OF ANHUI MEDICAL UNIVERSITY. The study period covered July 2020 to December 2021. Their serum tumor markers and CT features were collected. Low-dose CT scans have first detected the majority of SPNs for lung cancer screening in the hospital.
The inclusion criteria for this study were the following four: (1) age over 18 years old (2) patients with SPNs that measured 8 millimeters or smaller (3) all lesions are either lung cancer or benign lesions; and (4) they all have a definite diagnosis and have undergone serological testing.

*2.3. Variables.* The clinical characteristics extracted in our study includes: age, gender, Cyfra21-1, Neuron specific enolase (NSE), carbohydrate antigen 72-4 (CA72-4), stress corrosion cracking (SCC), precursor gastrin-releasing peptide (ProGRP), area and Carcinoembryonic Antigen (CEA). The age was categorized as <=49, 50-79 and >=80; the ProGRP was categorized as <=50, 50-150 and >=150; the NSE was categorized as <=15 and >15; the Cyfra21-1 was categorized as <=30 and >30; the CA72-4 was categorized as <=6 and >6; the SCC was categorized as <=2.5 and >2.5; the area was categorized as <=0.05 and >0.05; the CEA was categorized as <=5 and >5.

*2.4. Statistical Analysis.* Initially, univariate logistic regression analysis was used to assess the relationship between tumor markers and CT imaging features and SPNs patients. We then used multi-factor logistic regression analysis to identify risk predictors of lung cancer. Variables are included in the model when their p-value is less than 0.05. Variables are included in the model when their p-value is less than 0.05 Distinctness and correctness were used to evaluate the validity of the line graphs. In our study, we used the receiver-operator characteristic curve (AUC) to evaluate the ability of the model to accurately predict whether SPNs would transform into malignant tumors. The AUC is a metric that ranges from 0.5 to 1.0 and is used to assess the ability of a prediction model to distinguish between two groups. A higher AUC value indicates better discrimination, or the ability to correctly classify subjects into the appropriate group. A value of 0.5 represents a random chance of correctly discriminating the outcome, while a value of 1.0 indicates perfect discrimination. The calibration curve for the nomogram was generated through internal validation using 1000 bootstrap resampling. A wholly accurate nomogram will draw a calibration curve where a given group's observed and predicted probabilities would be along a 45-degree line. In all studies, p values less than 0.05 were considered statistically significant. The statistical analyses were performed by R 4.1.2.

## 3. Results

A total of 421 patients with SPNs that were 8 millimeters or smaller in diameter and were diagnosed as either benign or malignant were included in this study. The sample included 196 benign lesions and 225 malignant lesions. The nomogram model was developed and validated using these patients. The 421 patients included in the study were comprised of 297 men (70.1%) and 124 women (29.9%), with ages ranging from 20 to 90 years. Table 1 summarizes the tumor markers and CT imaging features of the patients. By univariate analysis, gender, age, ProGRP, CEA, Cyfra21-1, SCC, and nodal area were statistically different between lung cancer and benign lesions ($p < 0.05$). These variables were included in the multivariate logistic regression analysis to determine their relationship with the outcome of interest. The multivariate analysis indicated that the presence of lung cancer was significantly associated with ($p<0.001$), age (p (50-79) = 0.009, p (> =80) = 0.005), ProGRP (p (50-150) = 0.04, p (>150) = 0.004), CEA (p (>5) <0.001) and SCC (p (>2.5) =0.0016). However, Cyfra21.1 and nodule area were not statistically significant. This is detailed in Table 2.

There is ongoing debate and uncertainty surrounding the management of SPNs [13,14]. Although there are many mathematical models such as American Veterans Model (VA model) in clinical practice that can predict the possibility of the malignant tumor through the related symptoms of lung cancers, there are no researches of predicting the occurrence of malignant tumors through the combination of imaging features and tumor markers. In this study, a nomogram was developed to preoperatively identify SPNs based on CT imaging features and multiple tumor markers. The nomogram included five features: gender, age, ProGRP, CEA, and SCC. The calibration curve showed good agreement between the predicted probabilities and the actual observations with an internally validated AUC of 0.8474, as shown in Figure 1.

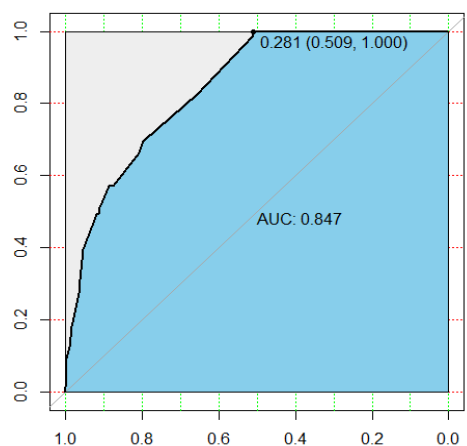

FIGURE 1: The ROC curve was used to evaluate the ability of SPNs to distinguish between the two groups.

Currently, there is no consensus on the surgical resection indications for SPNs that are 8 mm or less in diameter [15]. In order to conveniently provide effective reference and guidance for the clinical treatment and diagnosis of SPNs, we visualized the statistical model as a nomogram, which can quantify lung cancer risk, as shown in

Figure 2. In Figure 2, each variable's value is assigned a score ranging from 0 to 100. The scores for each variable are added together to give a total score. By mapping the total score to the lower point scale, the predicted probability of lung cancer can be determined by the corresponding cumulative sum. The calibration curve is illustrated in Figure 3. The point at the top right of the curve can be used to represent the distribution of the predicted probability of lung cancer. Falling on the 45-degree line can be indicated as a perfect prediction. The black dashed line indicates the cases where the predicted probability is exactly the same as the actual observation. The red line represents the entire cohort of 421 patients, while the green straight line indicates the bias corrected by 1000 replicate bootstrapping. These lines show the performance observed by the nomogram.

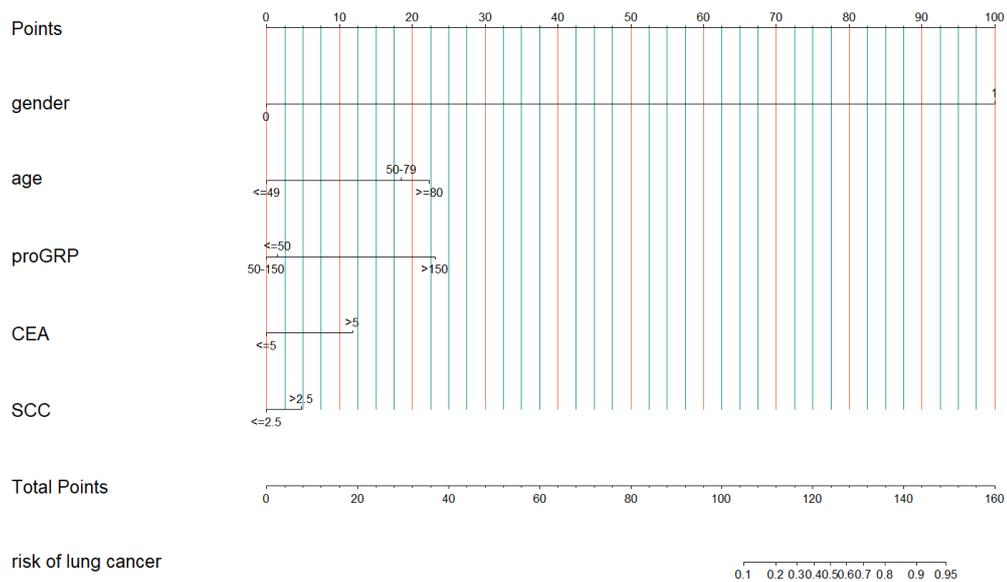

FIGURE 2: Nomogram for predicting the probability of lung cancer in patients with SPNs diameter of 8 mm or less.

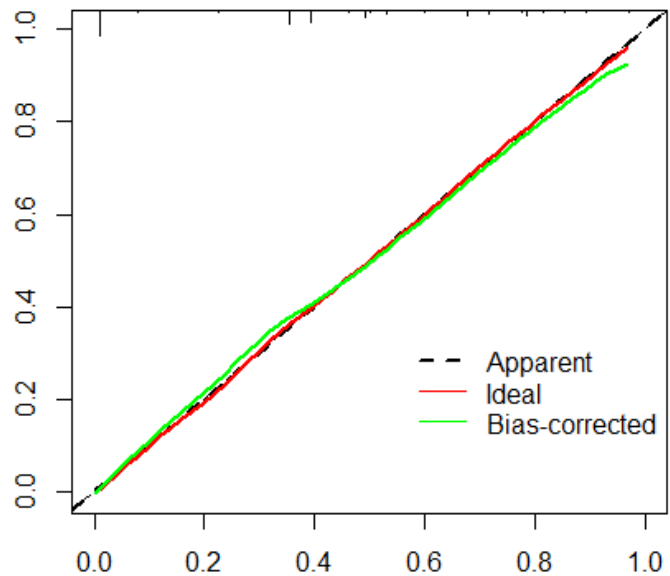

FIGURE 3: A nomogram calibration chart is internally verified.

*Table 1. Univariate analysis of tumor markers, imaging characteristics and risk of adenocarcinoma in lung cancer in 421 patients with SPNs less than 8 mm in diameter*

| Characteristic | training cohort | validation cohort | Overall | p-value[2] |
|---|---|---|---|---|
| **age** | | | | |
| <=49 | 17 (5.8%) | 10 (7.9%) | 27 (6.4%) | |
| >=80 | 24 (8.2%) | 11 (8.7%) | 35 (8.3%) | <0.001 |
| 50-79 | 253 (86.1%) | 106 (83.5%) | 359 (85.3%) | 0.002 |
| **gender** | | | | <0.001 |
| Mean (SD) | 0.690 (0.463) | 0.740 (0.440) | 0.705 (0.456) | |
| Median [Min, Max] | 1.00 [0, 1.00] | 1.00 [0, 1.00] | 1.00 [0, 1.00] | |
| **ProGRP** | | | | |
| <=50 | 109 (37.1%) | 60 (47.2%) | 169 (40.1%) | |
| >=150 | 12 (4.1%) | 6 (4.7%) | 18 (4.3%) | <0.001 |
| 50-150 | 173 (58.8%) | 61 (48.0%) | 234 (55.6%) | 0.3 |
| **NSE** | | | | |
| <=15 | 95 (32.3%) | 44 (34.6%) | 139 (33.0%) | |
| >15 | 199 (67.7%) | 83 (65.4%) | 282 (67.0%) | 0.092 |
| **Cyfra21-1** | | | | |
| <=30 | 276 (93.9%) | 119 (93.7%) | 395 (93.8%) | |
| >30 | 18 (6.1%) | 8 (6.3%) | 26 (6.2%) | 0.021 |
| **CA72-4** | | | | |
| <=6 | 243 (82.7%) | 102 (80.3%) | 345 (81.9%) | |
| >6 | 51 (17.3%) | 25 (19.7%) | 76 (18.1%) | >0.9 |
| **SCC** | | | | |
| <=2.5 | 215 (73.1%) | 94 (74.0%) | 309 (73.4%) | |
| >2.5 | 79 (26.9%) | 33 (26.0%) | 112 (26.6%) | <0.001 |
| **area** | | | | |
| <=0.05 | 207 (70.4%) | 84 (66.1%) | 291 (69.1%) | |
| >0.05 | 87 (29.6%) | 43 (33.9%) | 130 (30.9%) | 0.006 |
| **CEA** | | | | 0.8 |
| <=5 | 203 (69.0%) | 83 (65.4%) | 286 (67.9%) | |
| >5 | 91 (31.0%) | 44 (34.6%) | 135 (32.1% | <0.001 |

[1]n (%)    [2]Pearson's Chi-squared test; Fisher's exact test

*Table 2. Multivariate Logistic Regression Analysis of Lung Cancer Incidence Risk and Model Selection Factors*

| Characteristic | Univariate analysis | | | Multivariate analysis | | |
|---|---|---|---|---|---|---|
| | exp(Beta) | 95% CI[1] | p-value | exp(Beta) | 95% CI[1] | p-value |
| **gender** | 1.68 | 1.54-1.83 | <0.001 | gender | 1.49-1.81 | <0.001 |
| **age** | | | | | | |
| <=49 | — | — | | — | — | |
| >=80 | 1.60 | 1.26-2.03 | <0.001 | 1.36 | 1.10-1.67 | 0.005 |
| 50-79 | 1.35 | 1.12-1.62 | 0.003 | 1.24 | 1.06-1.46 | 0.009 |
| **ProGRP** | | | | | | |
| <=50 | — | — | | — | — | |
| >150 | 1.50 | 1.19, 1.89 | <0.001 | 1.34 | 1.10-1.63 | 0.004 |
| 50-150 | 1.05 | 0.96, 1.16 | 0.3 | 0.97 | 0.89-1.05 | 0.4 |
| **CEA** | | | | | | |
| <=5 | — | — | | — | — | |
| >5 | 1.24 | 1.12, 1.36 | <0.001 | 1.22 | 1.12-1.33 | <0.001 |
| **NSE** | | | | | | |
| <=15 | — | — | | | | |
| >15 | 1.09 | 0.99, 1.20 | 0.092 | | | |
| **Cyfra21-1** | | | | | | |
| <=30 | — | — | | — | — | |
| >30 | 1.25 | 1.04, 1.51 | 0.021 | 1.03 | 0.87-1.21 | 0.7 |
| **CA72-4** | | | | | | |
| <=6 | — | — | | | | |
| >6 | 1.00 | 0.89, 1.13 | >0.9 | | | |
| **SCC** | | | | | | |
| <=2.5 | — | — | | — | — | |
| >2.5 | 1.32 | 1.20, 1.46 | <0.001 | 1.12 | 1.02-1.23 | 0.016 |
| **area** | | | | | | |
| <=0.05 | — | — | | — | — | |
| >0.05 | 1.15 | 1.04, 1.27 | 0.006 | 0.94 | 0.86, 1.03 | 0.2 |

[1]CI = Confidence Interval

Reviewing the research of evaluating the malignancy of lung cancers, many mathematical models have been developed internationally, such as the Mayo model, which is the world's first predictive analysis variables model for assessing this kind of problem. The American Veterans Model, which is a more classic model but may not be make much sense for female patients since its data source was basically male [16]. Subsequently, Yang [17] use the logistic regression algorithm to build a model included 715 patients in five areas of China (Shanghai, Chongqing, Nanjing, Beijing, and Henan), with the AUC area 0.9151 in the training but 0.5836 in the validation set. There is also the PKUPH model, which was retrospectively analyzed in 371 patients and tested with an additional 145 patients with SPNs, predicted a sensitivity of 94.9% for benign and malignant tumors [18].

## 4. Discussion

Lung cancer used to occur mainly in men, so lung cancer is also called "male cancer", reports show that the number of lung cancer cases in women is currently growing faster than in men, but the overall incidence rate in men is still greater than in women [19]. According to some epidemiological surveys in recent years, the increase in the frequency of smoking among women is greater than that in men [20]. Women are more sensitive to cancer-causing tobacco compounds and are therefore more likely to develop lung cancer from exposure to tobacco [21]. This may be related to the higher estrogen levels in women, which promote the development of lung cancer [22]. The rapid growth of female smoking in recent years and the presence of long-term passive smoking in the family may be important reasons for the rapid growth of the incidence of lung cancer in women. These may be the cause of gender differences.

Age can cause long-term effects of external oncogenic factors in the body. Under the long-term stimulation of the body, the normal growth of cells is out of control, and then clonal proliferation appears, which eventually leads to the occurrence of tumors [23]. Therefore, with the increase of age, the effect of tumorigenic factors on the body will be led to a rise in the possibility of tumor occurrence, and lung cancer is no exception. Related studies have shown that the incidence of malignant tumors increases from the age of 40 and peaks at 80 [24].

As a non-invasive method, detecting tumor markers is the primary method of tumor screening. Tumor markers are biologically active substances that are abnormally synthesized and secreted by malignant tumor cells. They are often present in tumor patients' tissues, body fluids, and excreta [10,24,25]. Tumor markers are commonly used in the early diagnosis, detection and assessment of the efficacy of lung cancer. [26]. Its combination with CT can improve the specificity of SPNs diagnosis and be used as a supplement to CT for detection.

Neuroendocrine cells in the lungs contain ProGRP. The latest research suggests that cancer cells of small cell lung cancer (SCLC) can produce higher concentrations of GRP and ProGRP [27], and ProGRP is being gradually applied to the clinic as a newer tumor marker with higher sensitivity and specificity for SCLC diagnosis [28].

CEA is a glycoprotein secreted by cells of the gastrointestinal tract that can cause an immune response in the body. As a general rule, the content of CEA in patients with malignant tumors increases significantly. As a broad-spectrum tumor marker, CEA can be expressed in a variety of tumors. When the patients' condition improves, the content of CEA decreases, and when the condition worsens, it increases.

According to the literature, patients with lung cancer have significantly higher levels of SCC than healthy individuals [29]. SCC is more sensitive in diagnosing squamous cell carcinoma (SqCC) than other lung cancer, which is consistent with the conclusion that serum SCC is specific for SqCC in previous literature reports, therefore, in the clinic, SCC is often used to detect the therapeutic effect of SqCC, such as recurrence, metastasis, and prognosis [30]. Furthermore, SCC is not affected by gender, age, and whether or not to smoke [31].

Five meaningful risk factors were finalised for inclusion in a columnar chart predicting the probability of developing lung cancer. These risk factors were identified through the regression model as being important in predicting the likelihood of lung cancer. The total score was determined using the following five factors: gender, age, ProGRP, CEA, and SCC. Each factor has a defined value within a scale of 0 to 100. Predicting the probability of occurrence of lung cancer by using low scoring scales. The AUC of 0.8474 (95% confidence interval 0.8122-0.8825) was validated internally by nomogram, which has a good discriminatory power.

The management of SPNs remains controversial and uncertain. In this study, a nomogram was developed for preoperatively identifying lung cancer based on tumor markers and imaging features.

Our work can be improved in the following areas. First of all, we have selection bias. The patients with SPNs who were retrospectively reviewed in this study had a significant probability of having lung cancer at the time of hospitalization. This is because most SPNs of 8 mm diameter or smaller and SPNs that were repeatedly assessed using low-dose CT scan follow-up were excluded from the study. By using larger and more stable datasets, the nomogram prediction accuracy may be improved. There are two limitations to our model. First, the data used in this study was collected retrospectively. Second, the data set was insufficient in size. Internal validation of the model showed excellent discrimination and calibration, but the accuracy of using bar charts to predict the probability of lung cancer in a new patient population remains an important consideration. This is a significant concern when using a nomogram for predictive purposes. Further validity can be tested in the future by collecting additional data and externally validating the nomogram. This would help to confirm the rationality of the nomogram.

Another limitation of our model is that it was constructed using only CT imaging information and tumor markers such as NSE and CEA. It did not consider other factors that may affect the probability of lung cancer. To further improve the accuracy of the model and make it more suitable for use as a published nomogram, it would be helpful to have a more comprehensive recruitment process and to collect prospective data on factors such as family history and smoking of lung cancer. This would provide a more comprehensive view of the patient population and may help to improve the accuracy of the nomogram.

Finally, it is important to note that our data only included SPNs with a diameter of less than 8 mm and did not systematically study other types of lung cancer. In the future, we will also collect relevant data, conduct a comprehensive analysis of the status of lung cancers, and build a better predictive model related to them.

## 5. Conclusion

We found that patients with SPNs less than 8 mm in diameter may still evolve into lung cancer. However, there is no medical consensus on the indications for surgical resection of SPNs ≤8 mm in diameter [32]. Therefore, over management of these SPNS is very common in clinical practice.

Given the importance of timely surgical management decisions in the treatment of SPNs 8 mm or smaller, it is necessary to have an effective model to guide these decisions and prevent unnecessary treatment; although using frozen specimens can be a reliable and helpful diagnostic tool for evaluating SPNs and guiding clinical decisions about further patient management.

There is a need to develop a simpler, safer, and more accurate tool. Unlike previous studies that considered only CT imaging features, this study combined CT imaging features and tumor markers to create a statistical model that can accurately quantify the risk of lung cancer. The model may assist clinicians in making treatment decisions for SPNs of 8 mm or smaller.

In conclusion, the nomogram established can effectively predict the risk of lung cancer in patients with SPNs ≤8 mm in diameter. This nomogram is easy to use and has been validated through our study. By combining CT imaging features and tumor markers based on nomogram designed to assist in the decision-making process for these patients. The nomogram has high precision and demonstrates excellent discrimination and calibration. This means that it can be used by surgeons and clinicians to make more precise treatment decisions for individual patients.